\documentclass[useAMS,usenatbib]{mn2e}

\usepackage{graphicx}
\usepackage{inputenc}
\usepackage{pdflscape}
\usepackage{txfonts}
\usepackage[usenames]{color}

\def\mbh{$M_\mathrm{BH}$\/}

\def\lledd{$L/L_{\rm Edd}$}

\def\rfe{$R_{\rm FeII}$}

\def\feiiq{\rm Fe{\sc ii}$\lambda$4570\/}
\def\msol{M$_\odot$\/}

\def\rg{R$_{\rm g}$\/}
\def\chm{$c(\frac{1}{2})$\/}

\def\ltsima{$\; \buildrel < \over \sim \;$}
\def\ltsim{\lower.5ex\hbox{\ltsima}}  
\def\gtsima{$\; \buildrel > \over \sim \;$}

\def\gtsim{\lower.5ex\hbox{\gtsima}}

\def\civ{{\sc{Civ}}$\lambda$1549\/}

\def\civbc{{\sc{Civ}}$\lambda$1549$_{\rm BC}$\/}

\def\cm3{cm$^{-3}$\/}
\def\hb{{\sc{H}}$\beta$\/}
\def\hg{{\sc{H}}$\gamma$\/}
\def\hbbc{{\sc{H}}$\beta_{\rm BC}$\/}

\def\mgii{{Mg\sc{ii}}$\lambda$2800\/}

\def\niv{{\sc{Niv]}}$\lambda$1486\/}
\def\ciii{{\sc{Ciii]}}$\lambda$1909\/}
\def\oiiiopt{{\sc{[Oiii]}}$\lambda\lambda$4959,5007\/}

\def\oiiiuv{{\sc{Oiii]}}$\lambda$1663\/}

\def\siiii{Si{\sc iii]}$\lambda$1892\/}
\def\aliii{Al{\sc iii}$\lambda$1860\/}
\def\heiiuv{He{\sc{ii}}$\lambda$1640}

\def\feiiuv{{{\sc{Feii}}}$_{\rm UV}$\/}
\def\feiiopt{{Fe \sc{ii}}$_{\rm opt}$\/}
\def\feii{{Fe\sc{ii}}\/}
\def\siii{{Si\sc{ii}}$\lambda$1814\/}

\def\feiiil{{Fe\sc{iii}}$\lambda$1914\/}
\def\fe{{\sc{Fe}}\/}

\def\heiiopt{He{\sc{ii}}$\lambda$4686\/}
\def\fe76087{{\sc [Fe vii]}$\lambda$6087\/}
\def\oiii{{\sc [Oiii]}$\lambda$5007}

\def\kms{km~s$^{-1}$}

\def\rk{$R_{\rm K}$\/}
\def\ergss{erg s$^{-1}$\/}

\def\rk{{$R_\mathrm{K}$}\/}

\def\siiv{Si{\sc iv}$\lambda$1397\/}

\def\siiiuv{Si{\sc ii}$\lambda$1533\/}

\def\Gsoft{$\Gamma_\mathrm{soft}$}
\def\o4959{{\sc{[Oiii]}}$\lambda$4959\/}

\title{3C 57 as an Atypical Radio-Loud Quasar: Implications for
the Radio-Loud/Radio-Quiet Dichotomy}
\author[Sulentic et al. ]
{\parbox[]{6.5in}{ J.W. Sulentic$^{1}$\thanks{E-mail: sulentic@iaa.es},  M. A. Mart\'{i}nez-Carballo$^{1}$,
P. Marziani$^{2}$, A. del Olmo$^{1}$, G. M. Stirpe$^{3}$,  S. Zamfir$^{4}$, I. Plauchu-Frayn$^{1}$\thanks{Currently at Instituto de Astronom\'{i}a, UNAM, Ensenada, M\'{e}xico}}\\ \\
$^{1}${Instituto de Astrofis\'{\i}ca de Andaluc\'{\i}a, IAA-CSIC, Glorieta
de la Astronomia s/n, Granada, 18008, Spain}\\
$^{2}$INAF, Osservatorio Astronomico di Padova, vicolo dell' Osservatorio
5, Padova, 35122, Italy\\
$^{3}$INAF-Osservatorio Astronomico di Bologna, via Ranzani 1, Bologna,
40127, Italy\\
$^{4}$ Department of Physics \& Astronomy,University of Wisconsin, Stevens
Points, USA\\
}
\begin{document}
\date{}
\pagerange{\pageref{firstpage}--\pageref{lastpage}} \pubyear{2015}
\maketitle
\label{firstpage}
\begin{abstract}
Lobe-dominated radio-loud (LD RL) quasars occupy a restricted domain in the 
4D Eigenvector 1 (4DE1) parameter space which implies restricted geometry/physics/kinematics
for this subclass compared to the radio-quiet (RQ) majority of quasars. We discuss how this
restricted domain for the LD RL parent population supports the notion for a RQ-RL dichotomy
among Type 1 sources. 3C 57 is an atypical RL quasar that shows both uncertain radio morphology 
and falls in a region of 4DE1 space where RL quasars are rare.

We present new radio flux and optical spectroscopic measures
designed to verify its atypical optical/UV spectroscopic behaviour
and clarify its radio structure. The former data confirms that
3C 57 falls off the 4DE1 quasar ``main sequence" with both extreme
optical \feii\ emission (\rfe $\sim$ 1) and a large \civ\ profile
blueshift ($\sim$ -1500 \kms). These parameter values are
typical of extreme Population A sources which are almost always
RQ. New radio measures show no evidence for flux change over
a 50+ year timescale consistent with compact steep-spectrum
(CSS or young LD) over core-dominated morphology. 
 In the 4DE1 context where LD RL are usually low L/L$_{Edd}$ quasars we suggest that
3C 57 is an evolved RL quasar (i.e. large Black Hole mass) undergoing a
 major accretion event leading to a rejuvenation reflected by
strong \feii\ emission, perhaps indicating significant heavy metal
 enrichment, high bolometric luminosity for a low redshift source 
 and resultant unusually high Eddington ratio giving rise
to the atypical \civ.

\end{abstract}


\section{Introduction}
The origin of radio-loudness in quasars remains a perplexing question
fifty years after their discovery. Ironically radio-loud (RL) quasars were
the first to be discovered despite the fact that today they account for
only 8$\%$ of the low redshift quasar population. After 50+  years we do not know 
if RL quasars represent a distinct physical subset of the radio-quiet (RQ) 
quasar population or simply episodes through which all or most quasars pass. 
There is even confusion about the definition of a RL (or a RQ) quasar. In 
this paper we simplify the problem by focussing only on low redshift ($z \la$ 0.7) 
Type 1 AGN/quasars that   show broad (most in the range FWHM 
\hb\ =1000 -- 12000  \kms) emission line spectra including optical \feii\ 
emission. We assume that they represent the parent population of highly 
accreting AGN and interpret them in the 4DE1 context 
\citep[][hereafter Z08]{sulenticetal00a,sulenticetal07,zamfiretal08}.

We can unambiguously define a RL Type 1 quasar if we consider only the 
lobe-dominated (LD) RL sources which we assume to be the parent population 
of classical Type 1 RL quasars. They show radio/optical  flux ratios 
\rk $>$ 70, or better, $\log L_\mathrm{1415 MHz} > $ 31.6 \ergss\ Hz$^{-1}$. 
``Better'' because we avoid sensitivity  of optical flux measures to 
internal extinction/galaxy orientation effects. No bona fide LD sources 
are found below  our specified limits. Our  RQ-RL boundary is set by the 
radio luminosities of the weakest sources showing LD radio 
morphology \citep{sulenticetal03}. At low redshift (z \textless 0.7) LD
structure is seen only in quasars  with  bolometric luminosity brighter than 
log L$_\mathrm{bol}$= 44.0 \ergss. The 46 LD sources analysed by Z08 using an SDSS 
DR5 subsample of 470 quasars ($z <$0.7 and  brighter than g=17.5 or i=17.5) show a 
radio luminosity range of nearly 3 dex  ( $\log L_\mathrm{1415 MHz}$ =31.7-34.4  
\ergss\ Hz$^{-1}$) and also a 3dex optical range ($\log L_\mathrm{bol}=$44-47 \ergss).

Core-dominated (CD) radio sources cannot be used to define a RQ-RL boundary 
because they span a radio luminosity range of 6 dex ($ \log L_\mathrm{1415MHz}$ = 29 -- 35) 
from weak radio-detected RQ to the most luminous RL sources found in SDSS 
DR5 \citep[Z08 supplemented by][]{devriesetal06}. This corresponds to R$_K$ 
values from less than 10 to several thousand. The highest luminosity CD sources 
are interpreted as relativistically boosted LD sources oriented preferentially 
to our line of sight. Such sources often show apparent superluminal motions 
\citep[e.g.][]{zensusetal02}. As we proceed from the strongest towards weaker 
CD sources, and approach radio luminosities $ \log L_\mathrm{1415MHz} \sim$ 31.6 \ergss\ Hz$^{-1}$\ the problem 
becomes acute. They are not luminous enough to be aligned or misaligned LD 
sources. Across our adopted RQ-RL boundary (\rk =70) only CD (and weak core-jet) 
sources are found and in numbers increasing with decreasing radio power. A survey 
of radio emission for PG quasars \citep{kellermannetal89}
led to a suggested RQ - RL boundary near R$_K$ =10 using both LD and CD
detections. The choice of this boundary rather than \rk =70 
can have strong effects on statistical inferences when searching for differences between 
RL and RQ sources.

\begin{figure}
\centering
\includegraphics[width=\columnwidth]{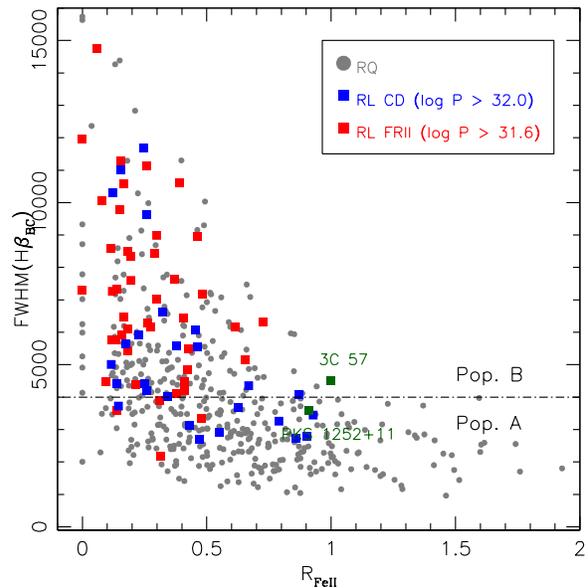}
\caption{Location of 3C 57 in the optical plane of the 4DE1 space defined by FWHM \hbbc\ vs. \rfe. 
Filled red squares are LD RL sources, filled blue square CD RL sources following the definition of 
\citet{zamfiretal08} with weaker CD RL ($\log L_\mathrm{1415MHz}<$ 32.0 \ergss\ Hz$^{-1}$)
omitted for clarity. Filled grey circles correspond to RQ  quasars.}
\label{Fig_E1plane}
\end{figure}

Contextualization can be helpful in relating quasar subclasses as well as
the relation of individual sources to specific subclasses. Towards this goal 
we adopted a 4D Eigenvector 1 parameter 
space \citep{sulenticetal00a,sulenticetal00b,sulenticetal07} based on four 
diagnostic measures: 1) Full width half maximum (FWHM) of broad ($\rm BC$) \hb;
2) Flux ratio of optical \feiiq\ blue blend
and broad \hb\ (\rfe); 3) profile shift at half maximum of high 
ionization \civ\, c(1/2) and 4) soft X-ray photon index (\Gsoft). In the 
4DE1 domain RL sources do not distribute like the RQ majority but instead show 
a preference for FWHM\hb\ $>$ 4000 \kms, \rfe $<$0.5, unshifted \civ\ profile 
and  absence of a soft Xray excess (i.e., $\Gamma_\mathrm{soft} \approx$2).  
This is especially true for the LD RL parent population. Figure \ref{Fig_E1plane} 
shows the 4DE1 optical plane for the SDSS DR5 quasar 
sample (Z08) with LD RL marked as filled red squares, luminous CD sources 
($\log L_\mathrm{1415MHz} >$ 32.0) as filled blue squares and  RQ quasars as filled grey circles.
We designate sources above and below FWHM \hbbc=4000\kms as Population B (Pop B) and Population A (Pop A)
respectively. The difference in 4DE1 domain occupation can be argued to be  
evidence that RL sources (the majority are Pop B) are fundamentally different 
from RQ quasars. However this interpretation is complicated by the fact that  
$\sim$ 40\%\ of RQ sources also occupy the same (Pop B) domain as the RL sources.

This paper considers an apparently nonconformist quasar 3C\ 57 (z= 0.67) that is 
unambiguously RL ($\log L_\mathrm{1415MHz} \approx 34.4$ \ergss Hz$^{-1}$ and 
$\log$ \rk $\ \sim\ $3.0).  As such we expect it to show  RL typical (average) 4DE1 
parameter measures: 1) FWHM \hbbc\ $\approx$ 6940 \kms; 2) \rfe\ $\approx$ 0.22; 3) 
\civ\ $c(\frac{1}{2})\approx$ +50 \kms\ and 4) \Gsoft\ $\approx$\ 2.15 
\citep{sulenticetal07}. In two of the four measures, 3C 57 is wildly discordant 
showing: \rfe $\approx$ 1 and \civ\ blueshift \chm$\ \approx -1500$ \kms\ 
\citep[][and this paper]{sulenticetal07}. These values are even extreme for RQ 
Pop A quasars. RQ sources have average values of \civ\ $c(\frac{1}{2})$\ 
=\ -580 \kms\ and \rfe\ $\approx$ 0.48 \citep{marzianietal96,sulenticetal07,richardsetal11}.

New spectra were obtained for 3C 57 with three motivations: 1) to verify
FWHM \hb\ and confirm the previous unusually high \rfe\ measures, 2) to search for
changes in the \hb\ profile that are sometimes observed in RL sources \citep{corbinsmith00} and
3) to obtain high S/N line profile measures of \mgii\ for use as a potentially
more reliable Black Hole (BH) mass estimator \citep{marzianietal13a,marzianietal13b}.
 New radio observations were also obtained to shed light on the ambiguous morphological interpretations
of 3C 57. It is very radio luminous and its apparent core-jet structure leads us to
expect flux variations over the 50+ year time span since its discovery.

The HST archive contained usable spectra for 130 low $z$\ quasars as of late 2006  
with a strong bias for RL  sources. Figure \ref{Figciv} shows a UV-optical plane of 4DE1
where FWHM normalized \civ\ centroid shift is plotted against \rfe . Open/filled black circles show 
RQ and RL Pop A quasars respectively. RQ and RL Pop B sources are indicated by open/filled red squares. 
Six of the 59 RL sources in the HST-FOS archival sample show a \civ\ blueshift 
$c(\frac{1}{2})$ larger than 1000 \kms while the mean centroid shift value for the RL sources 
is +52\kms \citep{sulenticetal07}. If one normalizes the shift value by FWHM \hbbc\ we 
find two sources that stand out: 3C 57 and PKS 1252+11. This plot best 
illustrates the nature of 3C 57. We present here new radio  and optical 
spectroscopic data that we hope will shed some light on the nature of 3C 57 (and PKS 1252+11)
and why it (they) contravenes the clear trends in 4DE1. UV data have been taken from
the HST-FOS archive and reanalysed. We use this to reconsider the definitions of radio-loudness.

{New optical and radio observations are presented in Section \ref{sec_obs} along with details of spectral 
analysis in \ref{sec_analysis}. Section \ref{results} presents results from analysis of new and literature data. 
Section \ref{discussion} discusses the relation between RL quasars, \civ\ blueshifts, and the possible RQ-RL
dichotomy. \ref{sec_wind} considers how winds might be affected by radio outbursts while 
section \ref{conclusions} summarizes our inferences and conclusions.

\begin{figure}
\centering
\includegraphics[width=0.72\columnwidth,angle=-90]{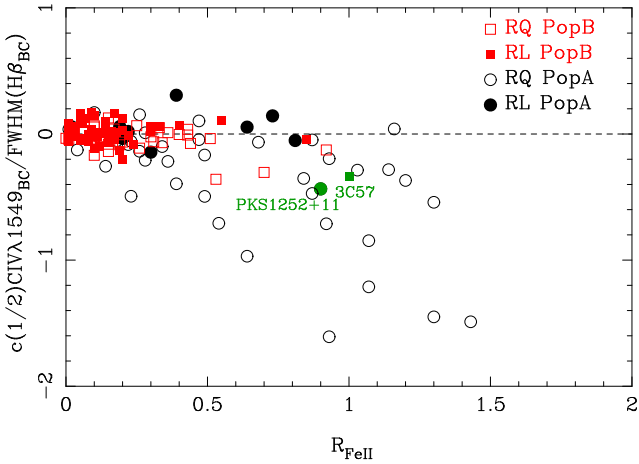}
\caption{Location of 3C 57 and PKS 1252+11 in a 4DE1 UV-optical plane defined by normalized FWHM \civ\ vs. \rfe. 
RL and RQ sources are marked as filled and open symbols respectively. 
Pop A sources are represented by black circles while Pop B by red squares.}
\label{Figciv}
\end{figure}

\begin{figure}
\centering
\includegraphics[width=\columnwidth]{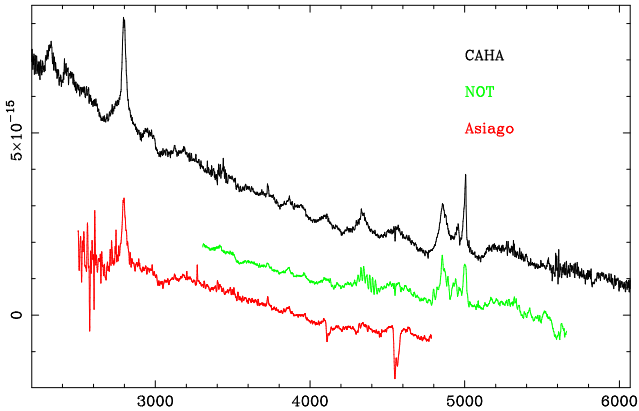}
\caption{New spectra for 3C 57 covering \mgii\ and \hb\ in the rest frame. 
Abscissa is wavelength in \AA\ and ordinate corresponds to  specific flux in units of \ergss\ cm$^{-2}$\ \AA$^{-1}$.
NOT and Asiago spectra have been vertically displaced by -1$\times10^{-15}$ \ergss\ cm$^{-2}$ and -2$\times10^{-15}$ \ergss\ cm$^{-2}$ respectively for presentation.}
\label{Fig_SpectraFull}
\end{figure}

\begin{table}
\caption{3C 57 Optical Observations}             
\label{tab_obs}      
\centering                         
\begin{tabular}{l c c c }        
\hline 
Obs:        &  ORM       & CAHA        &   Asiago    \\
\hline
Tel         & NOT2.5m       & 3.5m        &      1.82m  \\
Instr.      &   ALFOSC   & TWIN        &  AFOSC       \\
Scale    &  0.19"/px  & 0.56"/px    &  0.26"/px    \\
Grism       &   GR5      & T13 and T11   &   GR4         \\
Slit        &  1.3"      &  1.2"       &   1.26"     \\   
Disp   & 3.15\AA/px &  2.14\AA/px & 4.92\AA/px\\
            &            &  2.41\AA/px &            \\   
Range &5550-9400\AA & 3400-5800\AA & 3200-8200\AA \\
            &            &5450-10150\AA &               \\           
Date        & 29Aug2011  & 22Oct2012   &  05Dec2012  \\
Texp        & 4x900s     &  3x900s     &   3x1200s   \\
\hline                                   
\end{tabular}
\end{table}
\begin{table}
\caption{Radio Measurements at 5 GHz}
\label{tab_radio}      
\centering                         
\begin{tabular}{lll}        
\hline
  Flux (mJy)  &  Reference \\
\hline
\\
\multicolumn{2}{c}{3C 57}\\
\\
   1390 $\pm$80   &  \citealt{kuehretal81}\\
   1350           &  \citealt{wrightetal90}\\
   1372 $\pm$72   &  \citealt{griffithetal94}\\
   1441 $\pm$30   &  This paper (Observed in 2011)\\
\\
\multicolumn{2}{c}{PKS 1252+11}\\
\\
 1140 $\pm$50  &  \citealt{kuehretal81}\\
 1030 $\pm$70  &  \citealt{kuehretal81}\\
  641   &  VLA, \citealt{laurentmuehleisenetal97}\\
  961  $\pm$30  & This paper (Observed in 2011)\\
\hline
\end{tabular}
\end{table}

\section[observations]{Observations}
\label{sec_obs}

New long slit optical spectroscopic observations of 3C 57 were carried out 
in three different telescopes: Calar Alto Observatory (CAHA, Almer\' ia Spain), 
El Roque de los Muchachos Observatory (ORM La Palma, Spain) and Asiago Observatory (Italy),  
in three different runs within a 15 month period. Table \ref{tab_obs} summarizes the new data 
where we tabulate the instrumental setup for each observation as follows: 
telescope, spectrograph, spatial scale in arcsec/px, used grism, slit width 
in arcsec, spectral dispersion in \AA/pix and wavelength range. We also report 
the date of observation and the total integration time. The slit was oriented at 
parallactic angle to minimize effects of atmospheric differential refraction in 
the spectra. In the case of the TWIN spectrograph at CAHA with two arms, the 
observations were obtained simultaneously in the blue (grism \#T13) and red (\#T11) 
spectral regions. The three new spectra are plotted in Figure \ref{Fig_SpectraFull} 
with CAHA, NOT and Asiago spectra shown in black, green and red respectively. 
NOT and Asiago spectra have been vertically shifted to avoid confusion.

Data reduction was carried out in a standard way using the IRAF package. Spectra have 
been overscan corrected, nightly bias subtracted and flat-fielded with the normalized 
flat-field obtained after median combination of the flats. Wavelength calibration
was obtained using standard lamp exposures. The {\sc apall} task was used 
for object extraction and background subtraction. Instrumental response
and flux calibration were obtained each night through observations
of spectrophotometric standard stars from the list of \cite{oke1990} that we also 
use to remove telluric absorption bands.

New 5 GHz radio observations of 3C 57 were also carried out  during 
November 14, 2011 using the single dish 32 m IRA-INAF antenna at Medicina  
in on-the-fly cross-scan mode. PKS 1252+11, the other known RL with a large 
\civ\ blueshift, was also observed on Nov. 15 2011. Flux calibration was  
achieved using standard sources from the list of \citet{ottetal94}. Fluxes 
and uncertainties that include calibrator uncertainty are reported in
Table \ref{tab_radio}. 

UV data have been taken from the HST-FOS archive.
A re-analysis of UV spectra is presented in this paper. The 3C57 spectrum 
covers \siiv, \civ\ and \ciii\ and was presented in \citet{sulenticetal07}. 

\begin{figure}
\centering
\includegraphics[width=\columnwidth]{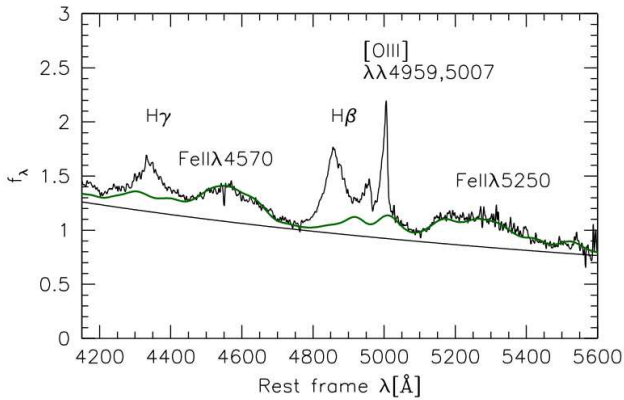}
\caption{{Continuum placement (filled line) for the normalised CAHA spectrum 
for the region of \hb. Abscissa is rest frame wavelength, ordinate is flux normalised at 5100 \AA. 
The dark-green line traces \feii\ emission. The strongest emission features are labelled.}}
\label{Fig_Fit}
\end{figure}

\subsection{Multicomponent $\chi^2$\ analysis}
\label{sec_analysis}

All strong emission lines in the spectra were fit using the IRAF task {\sc specfit} 
which employs a $\chi^{2}$\ minimization technique appropriate for nonlinear multicomponent 
analysis \citep{kriss94}. {\citet{marzianietal09} provide a thorough description of analysis procedures 
for the optical spectral range; \citet{marzianietal13a,marzianietal13b} of the \mgii\ range and 
\citet{negreteetal14} for the UV spectral range. 

All broad components were fit with Lorentzian 
profiles and the remaining lines with Gaussians (see \S \ref{ouprop} for a justification of 
this procedure). The {\sc specfit} task adopted a power-law continuum dominating over any host 
galaxy contribution (MgII absorption band was not detected), as well as \feii\ templates in both the 
optical and UV. We use the \feii\ templates obtained by \citet{marzianietal09} for the \hb\ spectral 
region and \citet{bruhweilerverner08} for the UV.  The {\sc specfit} task then 
scales and chooses an optical broadening factor for the template that minimises $\chi^2$. The $\chi^2$ 
minimisation procedures involve all components (continuum, \feii, emitting line components, etc.) 
simultaneously in order to construct a model of the spectrum. This allows us to 
consider the {\sc [Oiii]}$\lambda$4959 and\oiii\ lines with proper physical constraints (same 
shift and FWHM and a ratio 1:3; \citealt{dimitrijevicetal07}), and to avoid any subjectivity 
in the placement of continuum. Spectral coverage was wide enough to ensure that a portion of 
continuum with no or faint emission features was available to the fitting routine. Since we are 
mainly interested in a detailed reproduction of line profiles, a local continuum was fit for each 
spectral region considered in this study. Fig. \ref{Fig_Fit} shows the continuum and \feii\ placement from 
the {\sc specfit} analysis of the \hb\ spectral range  as an example of our approach. The fitting 
routine allows for the  pseudo continuum created by \feii\ emission and the fitted spectral range is wide 
enough to include wavelength intervals where the pseudo-continuum is low. Note that the continuum 
and \feii\ emission were not set a priori, but computed in the same minimum $\chi^2$ fits that allowed 
us to retrieve \hb\ and other emission line parameters. Line fits (after continuum subtraction) are shown 
in  Fig. \ref{Fig_Specfit}.} All fits were carried out over a wide wavelength range; however, 
we show four windows restricted to a lower range of the \hb\ (CAHA), \mgii\ (CAHA), \ciii\ 
blend(FOS) and \civ\ (FOS) centroids to facilitate comparison. Residuals are shown below each fit.

\begin{figure*}
\begin{center}
\includegraphics[width=0.69\columnwidth]{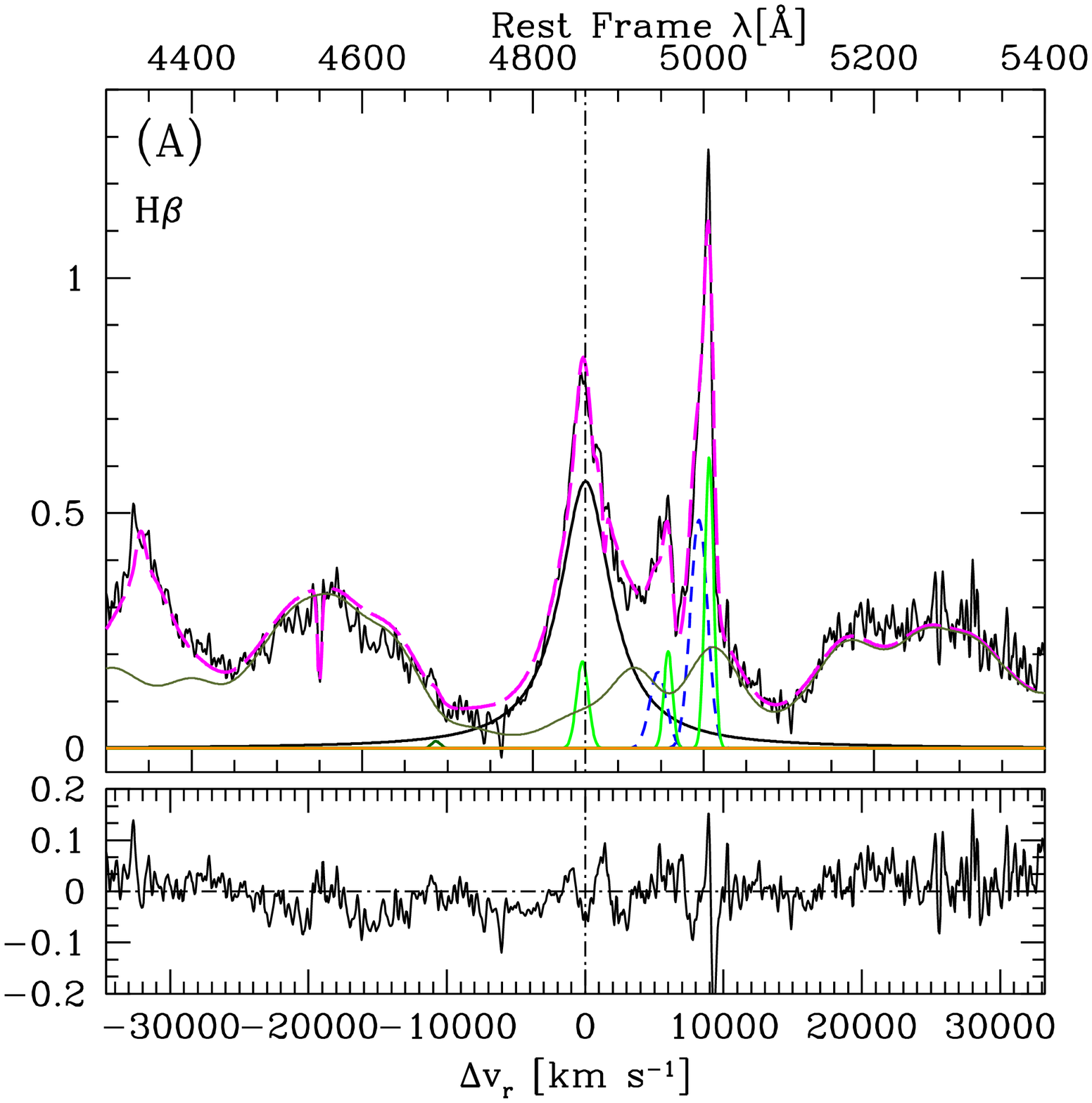}
\includegraphics[width=0.7\columnwidth]{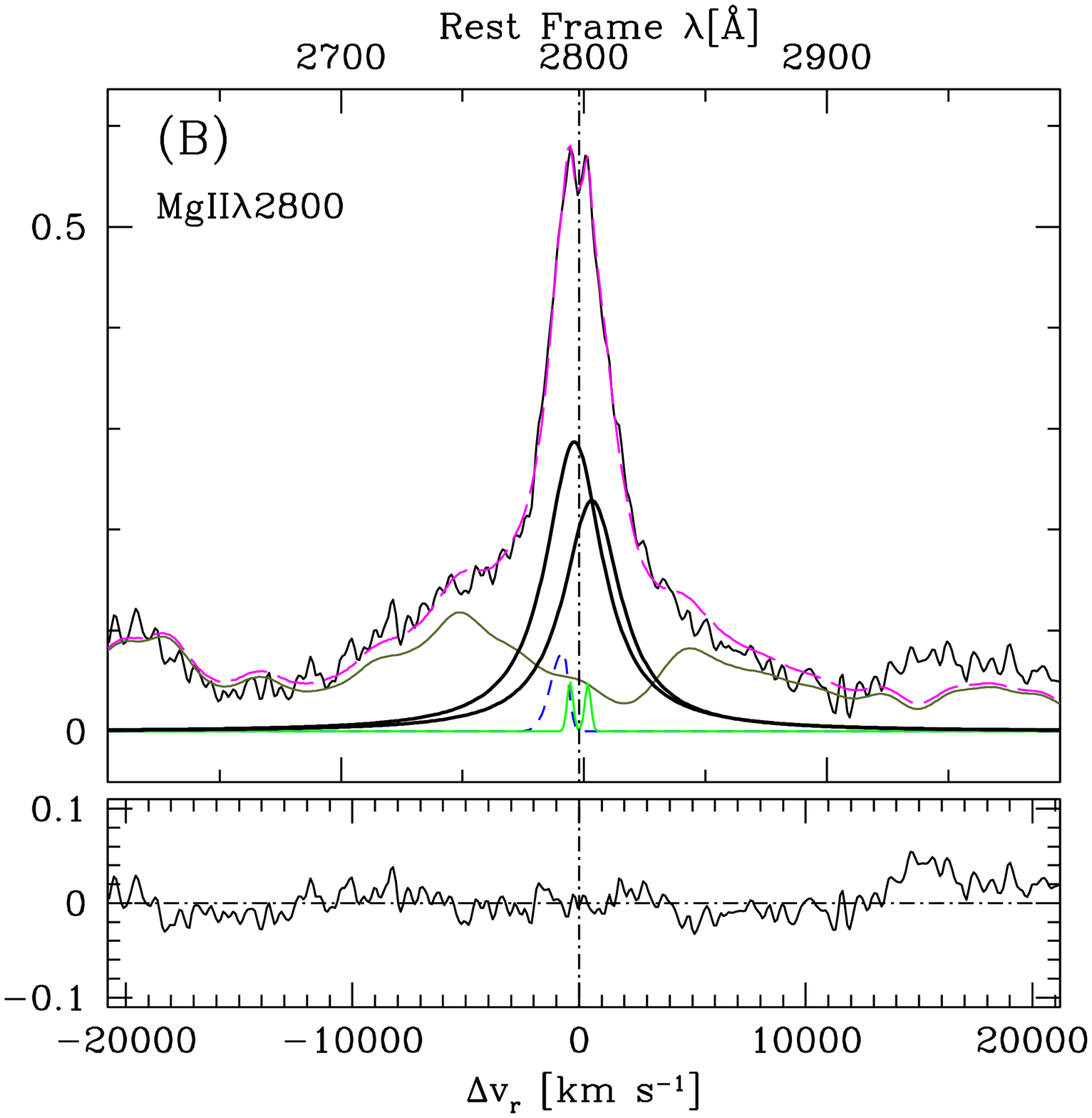}\\
\includegraphics[width=0.7\columnwidth]{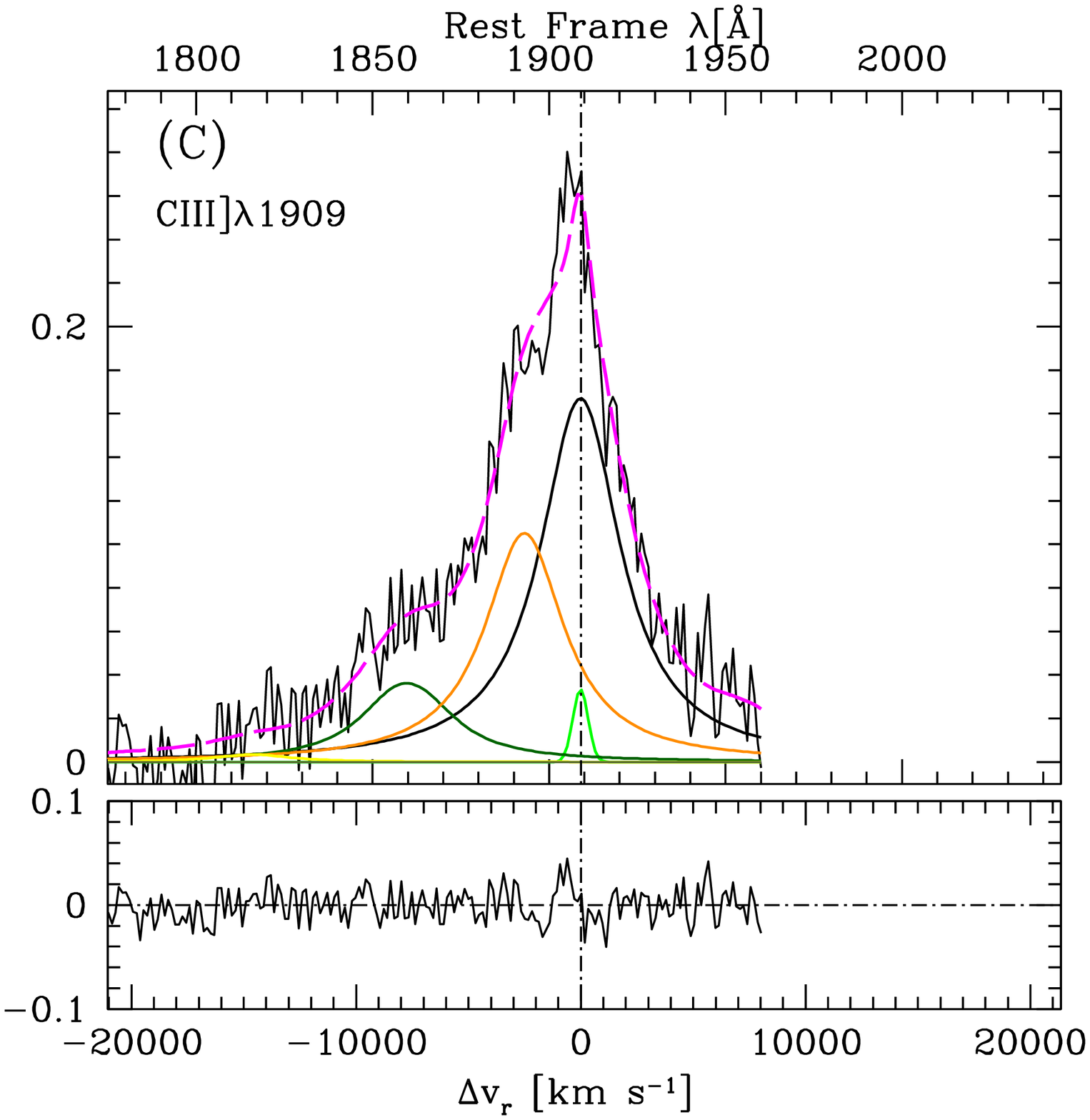}
\includegraphics[width=0.7\columnwidth]{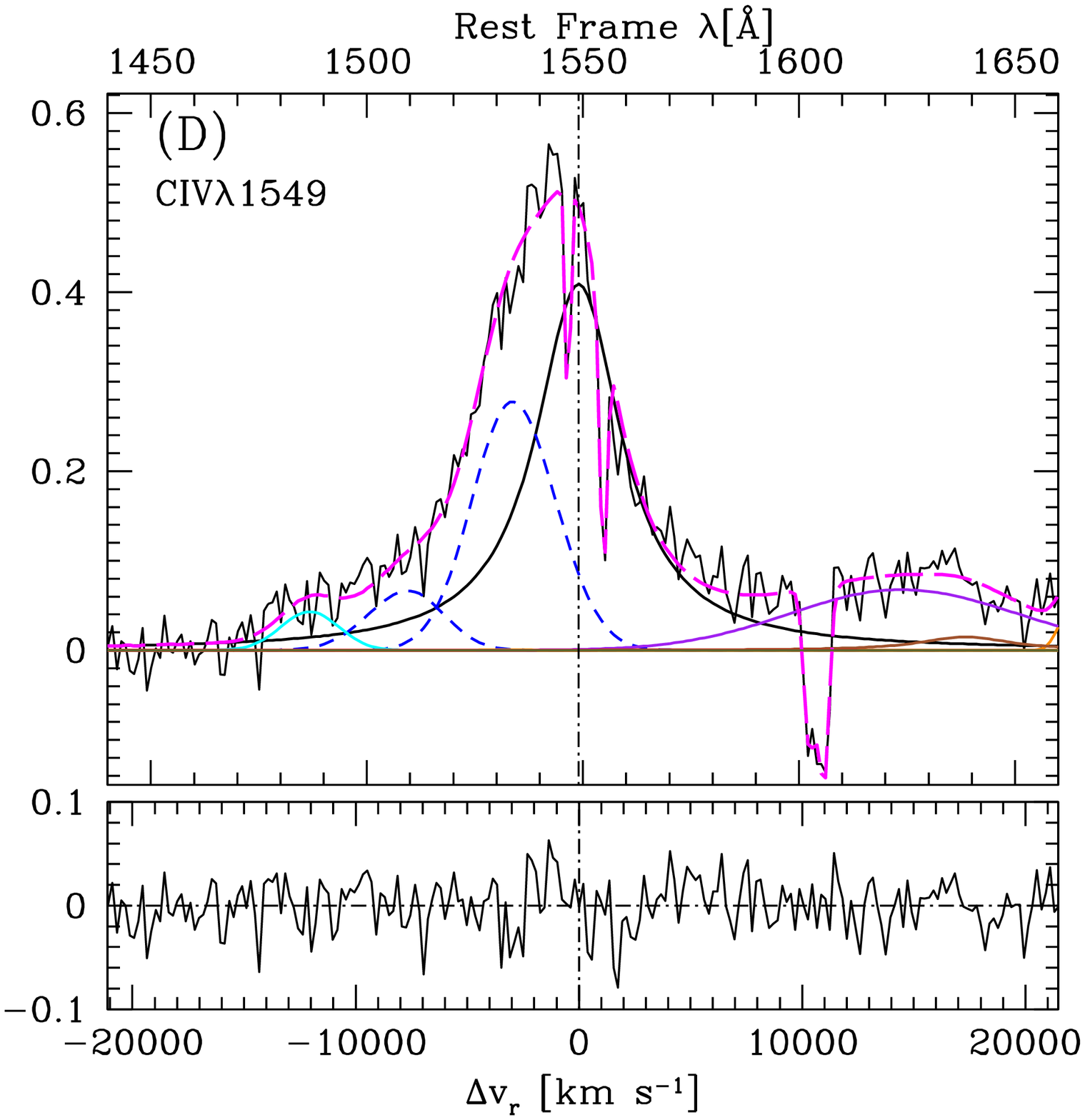}
\caption{Multicomponent fits for 3C 57. \hb\ and \mgii\ from CAHA spectrum and \ciii\ and \civ\ 
from HST spectrum. The upper abscissa is rest-frame wavelength in \AA, the lower abscissa is in radial 
velocity units, and the ordinate is specific flux per unit wavelength in arbitrary units.  The vertical 
long dashed line indicates the adopted rest frame. The black lines show the original continuum-subtracted 
spectra while the dashed magenta indicates the fit to the entire spectrum. The thick black and thin green lines 
show the broad and narrow components respectively. The blue-shifted  component is indicated by a dashed blue 
line when detected. The light grey lines trace \feiiuv\ and \feiiopt\ emission which is considered in all 
four fittings. In some of them (as \hb\ and \mgii) the contribution is very important but in others 
(as \ciii\ and \civ) it is negligible. In panel (C) apart of \ciii\ we show \siiii, \aliii\ and \siii\ 
in orange, dark green and yellow respectively. In panel (D) together with \civ\ we plot the \niv\ component
in cyan and the \heiiuv\ broad and blue component in brown and violet respectively.}
\label{Fig_Specfit}
\end{center}
\end{figure*}

We used a window from 4200\AA \ to 5500 \AA\ to fit \hb\ which include
\hg +{\sc{[Oiii]}}$\lambda$4363, \heiiopt\ and  \oiiiopt. Fits 
(see Fig.\ref{Fig_Specfit}a) assume that 
\o4959\ and \oiii\ have the same FWHM for narrow and blueshifted semibroad
components and that their flux ratio is \o4959\ /\oiii=1/3. Two components, 
narrow ($\rm NC$) and broad ($\rm BC$), were considered to model \hb. We also fit
\hg\ assuming the same number of components (narrow and 
broad) as \hb\ (and same FWHM values). From the \hb\ fits we obtain also an 
optical \feii\ flux (99$\times10^{-15}$ \ergss\ cm$^{-2}$) that together with measures 
of \hb\ (95$\times10^{-15}$ \ergss\ cm$^{-2}$) yields an estimate of 4DE1 parameter \rfe\ ($\sim$1).

Modelling of \mgii\ used a 2600\AA\ -- 3050\AA\ window where only this blend was detected. 
Each line of the \mgii\ doublet was fitted by  assuming broad, narrow  
and semi-broad blueshifted components (Fig \ref{Fig_Specfit}b).
The spectral window 1720\AA\ -- 1960\AA\ includes \ciii\, \siii\ (in yellow), \aliii\ 
(in dark green), \siiii\ (in orange) and \feiiil. FWHM of broad components of \siii, 
the doublet of \aliii, and \siiii\ are assumed to have the same value as \ciii. 
We also assume a narrow component for \ciii. For the sake of comparison 
we expand the window to 2045\AA\ in Fig \ref{Fig_Specfit}c.

The \civ\ model  (Fig.\ref{Fig_Specfit}d) includes unshifted and blue components 
(in black and blue respectively) and the fits to \niv\ (in cyan), \siiiuv\ , \heiiuv\  and \oiiiuv. 
We assume that \heiiuv\ also shows broad and blue components (shown in brown and violet respectively) 
with the same FWHMs and shifts  as broad and blue \civ. 
All these measurements are listed in Table \ref{tab_mess} where we include
measures of equivalent width, flux, FWHM and shift (with respect to the narrow component). We give broad, blue and total fitting parameters
for \civ. Reported uncertainties have been computed by measuring the effect, on each
parameter, of the change of continuum placement. To do that we repeat the {\sc specfit}
analysis changing the continuum by +/- 1 sigma level, that for the S/N accounted in
the spectra means 2-3\%. We also add quadratically the formal fitting error provided
by {\sc specfit}. The corresponding uncertainties are in agreement with those obtained for
similar S/N data by \cite{marzianietal03a,marzianietal13b}, \cite{negreteetal13,negreteetal14} and
 \cite{marzianisulentic14}. For the heavily blended lines at 1900\AA\, FWHM and
intensity uncertainties were estimated through a $\chi^2$ analysis of a mock
spectrum with the same S/N and intensity ratio and line width (as done in \citealt{negreteetal14}).

\section{Results}
\label{results}

\subsection{3C 57: A Steep Spectrum Radio-Loud Source}

3C 57 shows strong radio emission with $\log L_\mathrm{1415MHz}$ = 34.4 \ergss\  \citep{griffithetal94}
and \rk $\sim$1400 (see Table \ref{minisample}). Both parameters indicate a classical RL quasar. 
3C 57 has been a dangerous source in the  past because of its compact structure leading to 
inclusion in samples of CD sources \citep{willsetal92} while its steep spectrum (SS: $\alpha$\ =\ -0.7) 
warned that it was resolved \citep{morgantietal93}. More recent VLA maps \citep{reidetal99} finally
resolved the central core source into an elongated structure. The optical position 
of the quasar lies near the centre (see Fig \ref{position}) of the elongated source suggesting that it could be
an LD RL with lobe separation of $\sim$ 1.5 arcsec. If the two peaks are interpreted as lobes then 
the flux ratio suggests some degree of alignment towards our line of sight. If the quasar coincides 
with the weaker (north) peak then  a core-jet morphology would be implied putting it in the same 
class as PKS 1252+11.  The small separation (17 kpc projected) also suggests  either a preferred alignment 
and/or a young LD. A satellite component about 15 arcsec distant may be the relic of a past outburst. It 
shows no connection to the central elongated source which contains the bulk of the radio flux.  

\begin{figure}
\begin{center}
\includegraphics[angle=-90,width=0.7\columnwidth]{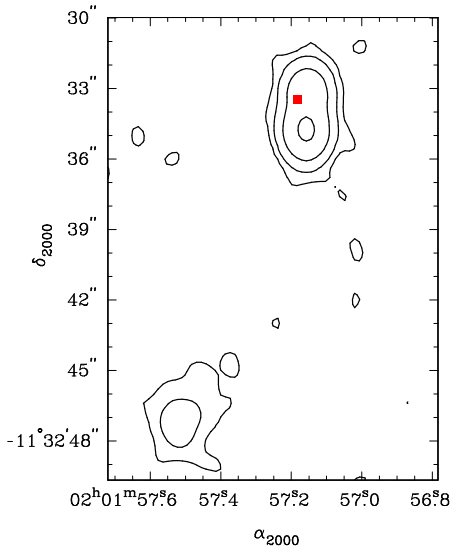}
\caption{Contour image of the VLA maps obtained by \citep{reidetal99} at 5 GHz with a resolution of 2". 
We plot levels 6$\times10^{-4}\times$ (1, 10, 100 1000) Jy.
The 3C 57 positions from the PPMXL catalogue is marked in red \citep{roeseretal10}. The optical/NIR position 
is consistent with being between the two radio blobs or closer to the fainter one.}
\label{position}
\end{center}
\end{figure}

3C 57 is not included in the 470 brightest SDSS-DR quasar sample because it lies 
outside the SDSS fields. It is useful to compare it with the $\sim$46 LD RL sources in the SDSS-DR5 
sample. Considering the distribution of all these sources in the plane defined by optical and radio 
luminosities ($\log L_\mathrm{bol}$ vs $\log L_\mathrm{1.4GHz}$; see Figure 6 in Z08) we find 
that 3C 57 is located in the upper right corner revealing extreme radio and optical properties 
relative to this local RL sample. Actually the bolometric luminosity of 3C 57 
($\log L_\mathrm{bol} \approx $  46.98) is  higher than all 470 sources in the SDSS-DR5 sample. 
In the upper corner of this plane we find the most luminous CD RL quasars which 
are the best candidates for relativistically  boosted sources while LD in this corner are the best 
candidates for young RL sources. Our motivation for new radio observations was to search for variability 
which might be expected if 3C 57 (or PKS 1252+11) were an aligned CD source. The 5 GHz specific fluxes 
reported in Table \ref{tab_radio} show  no evidence for changes in radio power over 30+ years. The 
high and stable radio power favours the idea  that 3C 57 is a young LD quasar (instead of a 
relativistically boosted source) oriented with jet lobe axis far from our line-of-sight.
Since the radio power did not change we can assume that the large radio power and the unusually 
large \civ\ blueshift coexist at the same time. Spectral index and linear size of 3C 57 meet defining 
criteria for  CSS sources \citep{odea98}. Absence of variability and resolved double radio morphology
are consistent with 3C 57 as a young somewhat aligned lobe-dominated source.

The new and older radio observations for PKS 1252+11 (Table \ref{tab_radio}) show evidence
for possible changes (fading) consistent with its Flat Spectrum (FS) CD morphology.

\begin{table}
\footnotesize
\caption{Optical and UV Measurements}             
\label{tab_mess}                            
\begin{tabular}{@{}l c c c c c@{}} 
\hline
Obs.                      & EW (\AA) & Flux$^a$ & FWHM$^b$  & Shift$^b$\\
\hline                    
\oiii                     & 8    & 13$\pm$1    &  725$\pm$57     &  +187   \\
\oiii$_{\rm BLUE}$        & 12   & 20$\pm$2    & 1426$\pm$145    &  -535 \\
\hb$_{\rm BC}$            & 62   & 95$\pm$10   & 4500$\pm$470    &  +223  \\ 
\feiiq                    & 56   & 99$\pm$10   &      ---        &  --- \\
\mgii$_{\rm BC}$          & 25   &122$\pm$10   & 3056$\pm$245$^c$& +189  \\
\ciii$_{\rm BC}$          & 11   &177$\pm$17   & 4318$\pm$430    & 0$^d$ \\
\siiii$_{\rm BC}$         &  7   &110$\pm$18   & 4318$\pm$430    & 0$^d$ \\
\aliii$_{\rm BC}$         &  3   & 39$\pm$12   & 4318$\pm$430$^c$& +248  \\
\civ$_{\rm  Tot}$         & 26   &567$\pm$57   & 6714$\pm$673    & -1454 \\
\civ$_{\rm BC}$           & 17   &376$\pm$32   & 4627$\pm$398    & +51   \\
\civ$_{\rm BLUE}$         &  9   &200$\pm$26   & 4729$\pm$602    &-3034   \\
\hline 
\end{tabular}
\hspace{20pt}
{\em $^{(a)}$} In units of $10^{-15}$ \ergss\ cm$^{-2}$. {\em $^{(b)}$} Relative to the 
central wavelength of the narrow component in \kms.{\em $^{(c)}$} FWHM of a single 
broad component. {\em $^{(d)}$} Fixed.
\end{table}

\subsection{Optical and UV properties}
\label{ouprop}

The new spectroscopic measurements confirm that 3C 57 shows mixed characteristics of 
a Pop A (\rfe $\approx$ 1) and Pop B  (FWHM H$\beta$ = 4500 \kms) source. Coming back 
to Figure \ref{Fig_E1plane} the marked CD RL are the sources most likely connected to the 
LD parent population in an orientation-unification scenario. If the radio axis 
is oriented perpendicular to a broad line emitting accretion disk in such a scenario, the 
CD tend to show smaller FWHM H$\beta$ and stronger \feii\ expected if the disk is oriented 
closer to face-on. 3C 57 is clearly an outlier in this plot. Technically it falls in 4DE1 
bin B2 \citep{sulenticetal02} but the bin boundaries shift upwards for higher luminosity 
quasars \citep{marzianietal09}. The bolometric luminosity of 3C 57 is high enough to place 
it in bin A2 which would be more consistent with measured \rfe\ and \civ\ shift values. 
We do not detect a significant \mgii\ blueshift which is also consistent with bin A2 quasars 
(\citealt{marzianietal13a}, blueshifts are most common in bin A3/A4 sources). The \civ\ 
profile shows an outflow-dominated profile that can be fit with: 1) a symmetric $\rm BC$ (assumed
equal to the scaled and shifted \hb\ profile) plus 2) a strong excess on the blue side -- modelled 
here using two gaussians.\footnote{Note that the total excess with respect to the $\rm BC$ is 
assumed to model the radial component of the wind/outflow. Its shape offers an observational 
constraint for outflow models. No meaning can be ascribed to the individual Gaussian components  
used to model the blue excess.} Our reanalysis of the UV spectrum (more detailed than the one carried 
out in 2007) confirms a large shift amplitude (typical of extreme Pop A sources) with total profile 
shift at FWHM $c(\frac{1}{2})=\ $ -1454 \kms and a shift of the modelled blue component of
$\approx$\ -3000 \kms (Table \ref{tab_mess}).

Another difference between Pop A and Pop B sources involves the shape of broad H$\beta$.
Pop A and Pop B profiles are best fit with Lorentzian and (double) Gaussian models respectively.
Given that the optical/UV data are consistent with a Pop A source we report in 
Table \ref{tab_mess} intensity values derived from Lorentzian fits. Normalized \civ\ blueshifts 
with amplitude greater than 1000 \kms are relatively frequent among largely RQ Pop A sources 
(Figure \ref{Figciv}). They are rare among RL sources not only at low-$z$\ but 
also at high $z$ for sources of similar luminosity \citep{sulenticetal14}.  Higher luminosity 
($\log L_\mathrm{bol}\geq$ 47) RL quasars no longer exist in the  local Universe. Results for high 
redshift quasars from an SDSS analysis suggest that \civ\ blueshifts may be much more common even 
among RL sources \citep{richardsetal11}. At lower $z$ or $L$, as can be seen in Figure \ref{Figciv}, 
there is a strong concentration of  mostly Pop B RL around 
mean values \rfe\ = 0.2 and  $c(\frac{1}{2})$ = 0 \kms. The RQ Pop A sources show a much 
wider parameter space dispersion towards larger\ \rfe\  and \civ\  blueshifts. 3C 57 in this plot 
(and PKS 1252+11) is located in the Pop A domain and shows the most extreme properties for a RL 
quasar in the HST low $z$\ sample.

3C 57 was monitored in the Catalina Real-time Transient Survey (CRTS,
\citealt{drakeetal12}) for almost seven years (from JD 53705.20724 --
56250.55267). No large V magnitude variations were detected with an
average V magnitude (computed from 269 observations)
$m_V = 16.061\pm0.09$. The rms scatter was 0.09 which is comparable
to the quoted uncertainty of individual measurements (0.08 - 0.1).
There is evidence for small amplitude  nonrandom variability
(possibly sinusoidal, amplitude 0.1) but a more detailed study would be needed.
There is no observational evidence of a significant change in the line 
profiles in our spectra. We verify this by scaling the NOT and CAHA spectra 
that cover \hb\ and the CAHA and Asiago spectra that cover \mgii. 
A difference between CAHA and NOT on the red side is due to the poor correction because of 
telluric absorption in the NOT. The Asiago spectrum is not well corrected for 
atmospheric extinction at its blue end. The line 
profiles look consistent if data quality is taken into account.
 
\subsection{RL Sources With Large \civ\ Blueshifts.}
\label{civblue}

3C 57 is not unique as a RL source  showing a significant \civ\ 
profile blueshift. It is one of six RL sources in the HST-FOS archive with \civ\
blueshift \ $c(\frac{1}{2}) < $ -1000 \kms\ \citep{sulenticetal07}. 
Observational properties are presented for these sources in Table \ref{minisample}. 
For each object we list the 4DE1 parameters (FWHM \hbbc, \rfe\, \civ\ shift, 
the normalized shift using FWHM \hbbc\ and \Gsoft), EW(\civ), accretion parameters 
($\log L_\mathrm{bol}$, log \mbh\ and log \lledd) and the radio 
properties ($\log$ \rk\, $\log P_{20cm}$, morphology, spectral index ($\alpha$) and lobe
separation). 
The 4DE1 parameters and EW(\civ) were taken from \cite{sulenticetal07}. 
In the case of 3C 57 we have recalculated FWHM \hbbc , \rfe , \civ\ shift and EW(\civ) using 
the new data. In the accretion parameters the bolometric luminosity ($\log L_\mathrm{bol}$) is estimated 
from the luminosity at 5100\AA\ assuming the standard bolometric correcting 
factor \citep[10;][]{richardsetal06}, 
the \mbh\ have been computed with the L-FWHM scaling from \cite{vestergaardpeterson06}.

\begin{table*}
\scriptsize
\caption{Object from \citet{sulenticetal07} with a \civ\ blueshifts $\le$ -1000 \kms}             
\label{minisample}      
\begin{center}                         
\begin{tabular}{lccccccccccccccr}        
\hline
Name            & \multicolumn{5}{c}{4DE1 parameters} & & \multicolumn{3}{c}{Accretion parameters}& & 
\multicolumn{5}{c}{Radio-loudness parameters}\\ 
\cline{2-6} \cline{8-10} \cline{12-16}
&    FWHM \hbbc  &  \rfe  & Shift & Shift$_\mathrm{Norm}$&  \Gsoft & EW(CIV) & $\log L$ & $\log $ \mbh & $\log $ \lledd & 
&$\log$\rk & $\log P$ & Morph.$^a$ & $\alpha$ & Sep.\\
 &    \kms  &    & \kms &  &    & \AA & \ergss &  \msol &  & & & \ergss Hz$^{-1}$ &  &  & arcsec\\

3C 057      &  4500	& 1.0	& -1454	& -0.32 & 2.28	&	18	&	46.83	&	8.98	&	-0.26	&   &3.078	&	33.71	&	SS LD/CJ	&	-0.58	&	1.50	    \\
Pictor A		& 18400	& 0.01	& -1110	& -0.06 & 2.34	&	176	&	43.72	&	8.51	&	-2.90	&	&4.217	&	32.60	&	SS LD	&	-1.58	&  480.00	\\
PKS 1252+11	&  3600	& 0.90	& -1570	& -0.44 & 1.88	&	23	&	46.85	&	8.99	&	-0.25	&	&2.971	&	33.56	&	FS CJ	&	-0.16	&	0.03	    \\
PKS 1355-41	&  8978	& 0.10	& -1070	& -0.12 & 1.96	&	74	&	46.13	&	9.17	&	-1.15	&	&2.856	&	33.34	&	SS LD	&	-0.82	&	48.00	\\
3C 390.3	    & 12688	& 0.12	& -1285	& -0.10 &1.8	&	132	&	44.83	&	8.78	&	-2.06	&	&3.450	&	32.48	&	SS LD	&	-0.78	&	210.00	\\
PKS 2349-01	&  5805	& 0.20	& -1170	& -0.20 &2.44	&	291	&	45.78	&	8.60	&	-0.94	&	&2.961	&	32.59	&	SS LD	&	-0.74	&	18.00	\\

\hline
\end{tabular}
\hspace{10pt}
{{\em $^{(a)}$} SS LD corresponds to Steep Spectrum Lobed Dominated and FS CJ 
corresponds to Flat Spectrum Core Jet}
\end{center}
\end{table*}

The radio parameters for 3C 57 and PKS 1252+11 have been updated to the values derived in 
this paper while those for the remaining objects have been calculated from the best 
available data in the literature (FIRST data preferred).

The sources in Table \ref{minisample} show a large diversity in 4DE1 optical/UV/X-ray measures as well as radio 
properties. Four of them fall in the Pop B optical domain of 4DE1 as expected for RL sources. 
The two associated with sources showing FWHM  \hb\ $>$ 10000 \kms (Pictor A and 3C 390.3) cannot 
be usefully compared with 3C 57 since they show such broad, and occasionally double-peaked, 
profiles. If we normalize the \civ\ shift by either FWHM \hbbc\  or FWHM \civbc\  the shift of such sources 
becomes much less significant compared to 3C 57. The normalized \civ\ shifts of the two other Pop B 
sources (PKS 1355-41 and PKS 2349-01) also become less prominent after 
normalization. The source that remains most similar to 3C 57 is PKS 1252+11 (Fig. \ref{Figciv}) 
which also shows strong \feii\ placing it outside of the Pop B domain.
Five of six sources in Table \ref{minisample} show steep spectrum radio SEDs and double
lobe morphology. The exception is PKS 1252+11 with a flat spectrum and core-jet
(CJ) morphology. CD sources come in two flavours. Those with steep (CSS)
and flat (FS/CS) radio SEDs. The former and latter are sometimes referred 
to as "young" and "frustrated" radio sources respectively.  The former  are interpreted
as birthing LD sources \citep[e.g.][]{vanbreugeletal84} while the latter might be failed 
attempts to generate LD sources, frustrated by a spiral host  galaxy 
morphology \citep{fantietal95} or perhaps an unfavourable BH spin.  3C 57 appears to be 
a good candidate for a young LD.  

Three of the sources in Table \ref{minisample} can be argued to show "young" radio
morphology -- two (3C 57 and PKS 2349-01) with very closely-spaced LD structure and PKS 1252+11 
interpreted as an LD precursor or dying frustrated source. The other three LD sources 
show wide enough lobe separations to preclude the assumption of a recent 
outburst. If we focus on 3C 57 and PKS 1252+11 as the sources with largest 
normalized \civ\ shift then we can argue that these sources should be
considered separately from the others. The outburst age might be
related with the surprisingly large \civ\ outflows where the RL activity
has not yet disrupted the wind. In such a scenario we assume that the scarcity of 
large \civ\ blueshifts in RL sources,  compared to RQ where they are common, involves 
quenching of the outflows by the onset of radio activity.

For the sake of comparison, one can also consider superluminal
3C 273 arguably the first quasar. It shows log(P$_\mathrm{20cm}$)$\backsim 34.31$ \ergss \ 
Hz$^{-1}$ similar to 3C 57 with a core-jet (or core-lobe) morphology similar 
to PKS 1252+11.  The \civ\ shift c($\frac{1}{2}$)=-552 \kms while \rfe\ = 0.57 and 
FWHM H$\beta\sim$3500 \kms places it in bin A2 of the 4DE1 optical plane. These properties 
make 3C 273 a Pop A  source consistent with the interpretation that is an aligned LD with 
undetected far-side lobe. In the optical plane of 4DE1 (Figure \ref{Fig_E1plane}) 
it lies at the transition region between LD and CD sources. One can search for additional 
RL sources showing large \civ\ shifts in the SDSS archive. The catalogue of \cite{shenetal11} 
reveals  (among 2347 RL quasars with \rk\ $>$ 100 and $z \ge 1.4$) a total of 43 sources 
with \civ\ blueshift $c(\frac{1}{2}) \ge $ 2000 \kms. Visual examination of the spectra
suggests that perhaps ten have high enough S/N to make the  blueshift credible 
and to tentatively assign them (using the UV criteria of 
\citealt{negreteetal14,marzianisulentic14}) to 4DE1 bins A2/A3. 3C 57 is not unique 
but apparently belongs to a tiny minority of RL sources showing significant 
\civ\ blueshifts and extreme (high \lledd) bin A2/A3 properties.

\section{Discussion} 
\label{discussion}

\subsection{\civ\ blueshifts and RL quasars.}

Low redshift quasar samples \citep{sulenticetal00a,marzianietal03a,zamfiretal10} 
show LD RL sources occupying a restricted zone in Eigenvector space (Figure 1). RL sources 
are largely what we call Pop B quasars which are characterised by: 
1) broad Balmer line profiles (FWHM \hb\ $>$\ 4000 \kms), 2) weak \feii\ optical emission
(\rfe\ $<$ 0.5), 3) absence of a \civ\ blueshift and 4) absence of a soft-X-ray excess.
 RQ sources sharing the Pop B zone with the RL also show weak or absent 
\civ\ blueshifts \citep{kuraszkiewiczetal04,sulenticetal07}.  Perhaps 
the absence of \civ\ blueshifts is related to Pop B  rather than to radio-loudness,  
or perhaps the RQ  Pop B sources are pre/post-cursors of RL activity. Do they all possess the 
same trigger (BH spin and/or host galaxy morphology) that enables radio-loudness?

There is of course also the issue of quasar luminosity. A recent comparison between 
low $z$\ sources (HST-FOS spectra) and 20 quasars at z $\sim$ 2.3\ $\pm$\ 0.2 using the GTC 
(in a narrow bolometric luminosity $\log L \approx 46.0\ \pm $\ 0.5 \ergss\ range 
\cite{sulenticetal14}) shows no evidence for significant  \civ\ blueshifts in either sample.  
Apparently \civ\  blueshifts are not common in Pop B sources below 
$\log L <$\ 46.5 \ergss\ and are possibly even rarer in RL sources. Is \civ\ blueshift 
correlated with source luminosity?  Our main study  
\citep[][]{sulenticetal07} -- that has the considerable advantage of reliable rest frame 
estimations based on narrow emission lines -- suggests that the answer is ``no". 
4DE1 parameters do not directly correlate with source luminosity which was found 
to be an Eigenvector 2 parameter \citep{borosongreen92}. All of our studies involving 
low $z$\ samples, and also a higher redshift VLT sample of 53 quasars \citep{marzianietal09}, 
point towards  {\bf Eddington ratio} as the principal driver of 4DE1 diversity and hence of
the \civ\ blueshift. 
The Pop A end of the 4DE1 main sequence shows systematically higher values of 
Eddington ratio than the Pop B end. Indeed we find  that 3C 57 shows an unusually high 
Eddington ratio for a RL quasar (Table \ref{minisample}).

Why do large SDSS samples \citep{richardsetal11} point toward a 
near ubiquity of \civ\ blueshifts including, albeit smaller, \civ\ blueshifts in the  
majority of RL quasars? Our mean/median \civ\ blueshift (-600$\pm$100 \kms) 
values for RQ sources is in general agreement with their SDSS results in 
the $\log L$ = 45-47 \ergss range. The systematic discrepancy appears in RL sources. 
The explanation can be due two effects: 1) they do not use the improved redshift 
determinations from \citet{hewettwild10} for RLs (only for RQ) and 2) they use the 
old definition for the RQ-RL boundary \rk\ $=$ 10. As explained earlier we adopt 
\rk $=$ 70 (and/or $\log L_\mathrm{1415MHz} = 31.6$ \ergss\ Hz$^{-1}$)  for this boundary.
In Figure \ref{Fig_E1plane} we adopt a more extreme boundary 
($\log L_\mathrm{1415MHz}=32.0$ \ergss\ Hz$^{-1}$) for RL CD sources only--under the assumption that 
marginally strong CD sources are unlikely to be aligned LDs. We argue that the pure CD 
population between \rk\ $=$ 10 and 70 are not classical RLs. Many may be LD precursors but 
they are still growing. There are a lot of sources in \citet{richardsetal11} with \rk\ 
between 10 and 70. If our interpretation is correct then we expect them to show 
\civ\ properties similar to the RQ  majority.  This would add a  large number of 
\civ\ blueshifts to the ``RL" population that do not belong there.
A new and larger SDSS based sample of LD sources \citep{kimballetal11} 
fully confirms our adopted $\log L_\mathrm{1415MHz}$\ RL limit.

\subsection{The nature of 3C 57 and implications for the RQ/RL Dichotomy.} 

The outstanding property of 3C 57  involves the coexistence of a large \civ\ 
blueshift with a RL source showing young core-jet (or aligned LD) structure. 
This implies that powerful relativistic ejection and a high ionization wind, thought to
be  associated with the accretion disk, can coexist. It also implies that powerful radio 
emission can occur in a quasar radiating at a relatively high Eddington ratio. This goes against 
previous arguments that powerful radio emission is unsteady, or even impossible 
at high \lledd\ \citep{fenderbelloni04,pontietal12,neilsenlee09}. Those theoretical 
arguments are however based mainly on  Galactic black hole X-ray binaries and it is not 
obvious that a close analogy with quasars can be made. It also appears to violate
4DE1 empiricism (Figure \ref{Fig_E1plane}) where RL sources occupy the low \lledd\
end of the quasar main sequence.

The two main emission  components in 3C 57 (at 5 GHz) show a projected separation 
of $\approx 17$ kpc\ which is approximately twice the effective radius ($R_\mathrm{e} \approx$ 8 kpc, 
\citealt{kotilainenfalomo00}) inferred for the host galaxy. The projected linear 
size is consistent with a CSS SED source showing kpc-scale radio emission
comparable to the size of the host galaxy. 
The development of extended (LD) radio emission requires a time  
$t\sim 3\times 10^7 (v_s/0.01c)^{-1}(R/\mathrm{100 Kpc}) \mathrm\, 
\mathrm{yr}$, where $v_s \sim 0.01c$ is the speed at the working surface of the jet 
\citep[][p. 298]{krolik99}, and $R$\ is the separation between the two lobes 
then  $t\sim 5 \times 10^6$ yrs in the case of 3C 57.   
The central black hole mass is estimated to be $\sim 10^9$ \msol\ (Table \ref{minisample}) with 
an efficiency of 0.07 \citep{netzer13}. 
The time needed for growing to this mass value is 
$\approx 6 \times 10^7$\ yr if the source has been constantly  accreting 
matter at a rate of $\approx$ 17 \msol yr$^{-1}$ -- implied by the observed 
$L$/\mbh\ ratio ($\log$\ \lledd $\approx$\ -0.26). The accretion time scale 
is an upper limit that includes the possibility of re-igniting the RL quasar activity.

\citet{wu09b} found that CSS sources exhibit a rather high median value of $\log$\lledd\ (\ =\ -0.56) 
that is typical of the A2 spectral region (-0.52) following  \cite{marzianietal13b}.
\citet{wu09a}  suggested that a relatively short duty-cycle is triggered by a 
radiation pressure instability within an optically thick, geometrically thin 
accretion disk \citep{czernyetal09}. In this interpretation, the detached 
radio component seen about 15 arcsec away from quasar may be a relic of past activity cycles.

The optical and UV spectra of 3C 57 are consistent with a scenario usually 
associated with Pop A sources of spectral types A2/A3 i.e., young or rejuvenated 
sources in the 4DE1 scheme. The wide majority of these 
sources are RQ and associated with higher accretion rates, 
enhanced star formation and chemical enrichment \citep{marzianietal01,sanietal10, 
marzianisulentic14}. These are the sources that most frequently 
show a  \civ\ blueshift in excess of -1000 \kms. \cite{zamanovetal02} suggest 
that \oiiiopt\ blueshifts are also associated with the high-ionization outflow originating in
these highly accreting sources. 3C 57 is not a blue outlier however, the \oiii\ profile shows a striking 
blueward asymmetry that can be modelled assuming a  core plus semibroad component 
with a relatively large blueshift. This is at variance with  evolved 
Pop B LD (FRII) sources that show large EW \oiii\ along with a narrow and 
symmetric core that appears to be consistent with dominance of the gravitational 
field of the host spheroid \citep[e.g.,][]{boroson03,marzianietal03b,marzianietal06,buttiglioneetal11}.
The relatively low equivalent width of \oiii\ as found in 3C 57 has also been associated with a relatively 
young, not fully developed narrow line region \citep{zamanovetal02,komossaetal08}. 

\subsection{Is the wind ubiquitous and how are outflow properties affected by 
radio-loudness? }
\label{sec_wind}

The jet kinetic power  can be written as 
\begin{equation}
L_\mathrm{j} =  P_\mathrm {j} \nu_\mathrm{j}\Omega_\mathrm{j} r^{2}
\end{equation}
where $ \Omega_\mathrm{j} r^2 = A_\mathrm{j}$ is the front end surface of the jet of 
solid angle $\Omega_\mathrm{j}$ at distance $r$\ from the black hole, $\nu_\mathrm{j}$ is the 
jet bulk expansion velocity and  $P_\mathrm {j}$\ the jet pressure within $\Omega_\mathrm{j}$.  
\begin{equation}
P_\mathrm{j}\approx 10^2 L_{j,44}\Omega_\mathrm{j,-3}^{-1}r^{-2}_\mathrm{0.1} \nu^{-1}_{j,5} \mathrm{dyne ~cm}^{-2}
\end{equation}
where   $L_\mathrm{j,44}$\ is in units of $10^{44}$ \ergss\ (typical RL values are
$\log L_\mathrm{j} > 44$, \citealt{guetal2009}), 
$ \Omega_\mathrm{j,-3}$\ is in units of $10^{-3}$\ sterad, $r_\mathrm{0.1}$ is $r$\ at 0.1
pc ($r_\mathrm{0.1pc} \approx 1000$ \rg\ for a black hole mass of $10^9$ \msol) and $\nu_{j,5}$\ 
is the jet bulk expansion velocity in units of $10^5$\kms. 
$P_\mathrm{j}$   exceeds the thermal pressure of the Broad Line Region (BLR) gas:
\begin{equation}
P = \frac{\rho k
T}{\mu m_\mathrm{p}}   = {n   kT}    \approx 1.38\, 10^{-3} n_9 T_4 ~\mathrm{dyne
~cm}^{-2}
\end{equation}
(where $\rho$\ is the mass density, $T_4$\ the temperature in units of
$10^4$K, and $n_9$\ the number density in units of $10^9$ \cm3), as well as the
hydrostatic pressure of a column of gas like the one expected to emit 
the blueshifted component ascribed to the accretion disk wind:
\begin{equation}
P_\mathrm{hyd} \approx \mu n m_\mathrm{p} h \frac{GM}{r^2} \approx  
\mu m_\mathrm{p} N_\mathrm{c} \frac{GM}{r^2} \approx  
0.03 \, N_{c,22} M_9 r_{0.1}^{-2}.
\end{equation}
where $N_\mathrm{c,22}$ is ambient gas column density in 
units of $10^{22}$ cm$^{-2}$, and $M_9$\ is the black hole mass in units of $10^9$\ solar masses. 
The first implication is that there 
should be a zone of avoidance close to the radio axis. A second 
implication is that the cocoon associated with the powerful relativistic 
ejection is also expected to sweep the gas within the broad line region. 
Elementary considerations based on the model of \cite{begelmancioffi89}
would suggest a cocoon  pressure in  directions perpendicular to that of the jet 
propagation:
\begin{equation}
P_\mathrm{c} \approx   {(P_\mathrm{j}\rho v_{j}A_{j})^{\frac{1}{2}}}/{\pi r_\mathrm{c}^2} 
\approx 3\, {(L_\mathrm{j,44} n_{9} v_\mathrm{j,5} \Omega_\mathrm{j,-3})^{\frac{1}{2}}} r_\mathrm{c,0.1}^{-2}
~\mathrm{dyne ~cm}^{-2}. 
\end{equation}
Therefore, if the cocoon side pressure is as strong as
inferred from these elementary computations (that neglects general relativistic and
magnetohydrodynamical effects  associated with the  jet tight   collimation),  we
expect a strong, destructive effect on a high ionization wind, especially in the
innermost BLR.

Observations of powerful RL sources reported  in \S \ref{civblue} indicate that the
high-ionization  outflow producing \civ\ and other lines is not suppressed, even if
hampered or altered.   There is a  different dependence on luminosity of the  median
and average \civ\ shift  in the \citet{shenetal11} data for both RQ and RL: for RL, 
\civ\ shifts are smaller amplitude and the luminosity dependence is shallower, 
with shifts above $-1000$ \kms\ being very rare for RL sources
\citep[c.f.][]{richardsetal11}.

A first inference could be that RL activity produces a 
wider cone of avoidance around the disk axis: i.e. suppresses emission 
along radial lines of flow close to the jet axis. In this case the outflow 
may be more equatorially confined giving rise to more symmetric profiles 
and to systematically lower shifts for RL, especially for CD RLs where 
the flow should be viewed pole on.  It is not clear whether this is the 
case: the sample of \citet{sulenticetal07} is small, while 
\citet{richardsetal11} include many core sources that may not be RL
(\rk=10-70). In this interpretation the radio morphology of 3C 57 may hint at a favourable 
orientation, coupled  with a significant outflow due to  the high \lledd. 
Another possibility (not conflicting with the previous one) is that the 
wind is  forced to start at larger distances from the central black hole 
(the cocoon pressure decreases with $r^{-2}$) reaching  a lower terminal 
velocity (still significantly above the escape velocity from the system). 
Larger emitting distances may also be  consistent with the models
of \citet{zamanovetal02} and \citet{komossaetal08}. 

Large (above 1000 \kms) \civ\ blueshifts appear to be rare: the tentative 
estimates of \S \ref{civblue} yield a prevalence 0.5 \%\ -- 2\%. The rarity 
of these sources suggests another possible explanation for the 3C 57 (and PKS 1252+11) 
blueshift: the radio activity  ignited too recently to have yet disrupted the wind. For 
3C 57 ($\log$ \mbh $\approx 9$) the dynamical timescale of the BLR is $\sim$ 100 yr. The 
derived  \mbh is large this indicates that 3C 57 is a rejuvenated quasar. If the 
duty cycle of   rejuvenated quasars is $ \ltsim 10^4$ yr  \citep{czernyetal09}, then
one may expect $\sim 1$\%\ of sources whose BLR has not been yet fully affected 
by the onset of radio activity.

\section{Conclusions}
\label{conclusions}

The 4D Eigenvector formalism reveals that the majority of RL quasars show
a restricted zone of parameter space occupation compared to the RQ majority. This
restriction is clearest when we focus on the unambiguously RL LD sources. They show 
restricted ranges of radio power ($\log L_\mathrm{1415 MHz}> 31.6$ \ergss Hz$^{-1}$), 
bolometric luminosity ($\log L_\mathrm{bol} > 44.0$ \ergss), FWHM \hb\ ($>$ 4000 \kms), 
\rfe ($<$\ 0.5), \Gsoft ($<$\ 2.5) and modest \civ\ blueshifts. Since the defining 4DE1 parameters 
are assumed to measure aspects of BLR physics and source geometry/kinematics this implies 
either: 1) If all quasars are capable of radio-loudness then important physical and/or kinematic 
properties of the BLR must change before the onset of a RL event  or, 2) as an alternative  
reflected in the Pop A- Pop B distinction, RL represent a distinct class of quasars driven 
perhaps by different BH spin and/or host galaxy morphology. The RQ sources sharing the same 
4DE1 parameter domain with the RL might represent currently radio-inactive Pop B quasars. 
We see a RQ-RL dichotomy if we consider only LD sources (as the RL parent population). CD sources 
are likely a mix of (rare) aligned LD sources, LD precursors and frustrated cores incapable of
producing classical LD structure. CD sources above $\log L_\mathrm{1415 MHz}< 32-32.5$ \ergss\ Hz$^{-1}$\ 
distribute in 4DE1 as expected if they are aligned LD (e.g. FWHM \hb\ near lower limit of LD) in an
orientation unification scenario while CD with   $\log L_\mathrm{1415 MHz}<32-32.5$\ do not. 
The weaker CD sources also distribute in 4DE1 space the same as the RQ majority.

There are always exceptions to the rule. The cases of 3C 57 and PKS 1211+11 show that 
a prominent high-ionization outflow probably 
driven by radiation pressure can coexist with powerful radio emission, although the 
simultaneous detection of both phenomena appears to be rare. This result suggests that high 
accretion and relativistic radio ejection may not be mutually exclusive for supermassive black holes, 
as found in the case of stellar mass black holes, and, at the same time, that radio emission has 
a quantitative effect on the high-ionization outflows.

3C 57 shows extreme optical and radio properties compared to the local RL
population and is therefore unambiguously RL ($\log L_\mathrm{1415MHz} 
\approx 34.4$ \ergss\ Hz$^{-1}$\ and log \rk\ $\sim$ 3). However it shows two 4DE1 
parameters that are highly discordant with the RL majority: unusually 
strong optical \feii\ emission (\rfe\ $\sim$ 1) and a large \civ\ blueshift 
$\sim$ -1500 \kms. It also shows an estimated Eddington ratio ($\log$ \lledd$\approx -0.26$) 
much higher than the majority of RL quasars and typical of a Population A2 source. 
VLA maps resolve it leading to an interpretation of 3C 57, which shows a CSS radio SED, 
as a core-jet or aligned LD source. The radio flux stability favours a young LD quasar. 
The \civ\ profile blueshift implies that there is a wind or outflow from a highly accreting 
disk. The general absence of \civ\ blueshifts in RL sources suggested the onset of radio 
activity somehow disrupts or confines the wind. A search of the SDSS quasar catalogue
suggests that 3C 57 belongs to a tiny minority of RL sources with significant \civ\
blueshift and high Eddington ratio.

It is clear that whatever the physical properties of the BLR
in normal quasars, the RL show a restricted range in those
properties presumably connected to their large \mbh\ and
low \lledd . 3C 57 is then likely hosted by an early-type galaxy
if the large BH mass implies a large bulge mass via
the  BH mass -- bulge mass correlation. The unusual properties are
most easily understood if 3C 57 is undergoing an apparently rare major
accretion event. This assumes that the rare unusually strong FeII emission
in a RL is a signature of such events. This causes 3C 57 to show properties
typical of the opposite end of the 4DE1 main sequence (higher BLR density, 
metallicity and accretion disk wind). The CIV wind is either too strong to be disrupted
or the event is so recent that this disruption has not yet occurred.

\section*{Acknowledgements}
We acknowledge Dra Simona Righini for observations in INAF-IRA radiotelescope station in Medicina (Italy).
We would like to thank Drs. Jaime Perea and Isabel M\'{a}rquez for all the fruitful discussions on the 
subject and their help with the observations. We also thank the anonymous referee for 
many useful comments which helped to significantly improve the presentation of our analysis. 
Part of this work was supported by Junta de Andaluc\'{i}a through Grant TIC-114 and 
by the Spanish Ministry for Science and Innovation through 
Grants AYA2010-15169, AYA2011-1544-E and AYA2013-42227-P. This research is based in part on data 
obtained with the 1.82m Copernico Telescope at the Asiago Observatory.  Based partially on 
observations made with the 3.5m telescope at the Spanish-German Observatory in Calar Alto 
(CAHA, Almer\'{i}­a Spain) jointly operated by the 
Max-Planck-Institut f\"{u}r Astronomie Heidelberg and the Instituto de Astrof\'{i}­sica de 
Andaluc\'{i}­a (CSIC). We thank all the CAHA staff for their high professionalism and 
support with the observations. Some data presented here were obtained with ALFOSC, which is 
provided by the Instituto de Astrof\'{i}sica de Andaluc\'{i}a (IAA) under a joint agreement 
with the University of Copenhagen and NOTSA.

\bsp

\label{lastpage}

\end{document}